\documentstyle[12pt, fleqn]{article}
\begin{document}

\title{\bf Relativistic Calculations for Photonuclear
           Reactions \\
           (II). Nonrelativistic reductions
	   and nuclear medium effects\thanks{
           Work supported in part by the Natural Sciences and
           Engineering Research Council of Canada} }

\author{\bf  M. Hedayati-Poor, J.I. Johansson and H.S. Sherif  \\
             Department of Physics, University of Alberta \\
             Edmonton,  Alberta, Canada T6G 2J1}

\date{\today}

\maketitle

\begin{abstract}

The relativistic amplitude for the direct knock-out contribution
 to $\left( \gamma, p \right)$ reactions on nuclei
is reduced to a nonrelativistic form using an
effective Pauli reduction scheme. The reduction is
carried out to second order in the inverse nucleon mass.
It is found that the interaction Hamiltonian appearing in the
nonrelativistic amplitude has significant dependence,
starting at second order, on the vector and scalar mean
nuclear potentials.
These strong medium modifications are absent in traditional
nonrelativistic calculations.
Detailed comparisons show that these modifications are crucial
to understanding the differences between relativistic and
nonrelativistic models.
These differences are also examined through reduction of the
relativistic amplitude via the Foldy-Wouthuysen
transformation.
Similar medium modifications are obtained in this case as well.
We discuss the implications of these medium
modifications for the consistency of existing nonrelativistic
calculations.

\end{abstract}

\section*{Introduction}

Recently  Lotz and Sherif have calculated the relativistic amplitude
for the direct knock-out contribution to $\left(\gamma,p\right)$
reactions in a distorted wave approximation \cite{LS88}.
Similar calculations have also been given by McDermott {\em et al.}
\cite{Mc88}.
The results obtained from these relativistic calculations were found to
be in better agreement with experimental data than those of nonrelativistic
distorted wave Born approximation (DWBA) calculations \cite{TU85}.
The latter calculations are based on the use of Schr\"{o}dinger
wave functions to describe the bound and continuum nucleons.
The nuclear current is obtained through a Foldy-Wouthuysen
(FW) transformation of the relativistic Hamiltonian describing
the interaction of a photon with a free nucleon \cite{FW,MVVH}.

In this paper we clarify the main differences
between the relativistic and nonrelativistic calculations.
We obtain the nonrelativistic amplitude from the relativistic
distorted wave S matrix through two different reduction schemes.
The first of these is the Pauli reduction in which the relativistic
S matrix of the $(\gamma,p)$ reaction is expressed in a form that
involves Schr\"{o}dinger-like wave functions and an effective
interaction Hamiltonian.
This Hamiltonian is expanded in
powers of $1/\left( E + M \right)$, where $M$ is the nucleon mass
and $E$ is its total energy.
The first order terms in this expansion,
in the limit $E \rightarrow M$, reproduce
the form of the usual nonrelativistic amplitude.
A characteristic feature of the higher order terms is the dependence
of the interaction Hamiltonian on the nuclear vector and scalar
potentials.
A short account of this discussion has been published \cite{HS94}.
This scheme has also been applied to a similar study of the
relativistic and nonrelativistic descriptions of $(e,e'p)$
reactions \cite{HJS94}.

The second approach is based on the use of the Foldy-Wouthuysen
transformation.
The relativistic knock-out amplitude is written for a model case
in which the initial bound and final continuum nucleons are
described by a single Dirac Hamiltonian with real vector and
scalar potentials.
Through the FW reduction we are able to write a nonrelativistic
limit of the amplitude to various orders in the inverse nucleon mass.
Again we find that the interaction Hamiltonian is dependent on the
strong nuclear potentials.

We begin section \ref{relativistic} by summarizing the calculation
of the relativistic S matrix describing the direct knock-out
contribution to $(\gamma,p)$ reactions on nuclei.
In addition we provide some discussion of the Dirac
equation containing strong scalar and vector potentials.
In section \ref{Pauli} we illustrate the formalism of the
Pauli reduction and show how the relativistic S matrix
of the $(\gamma,p)$ reaction is reduced to an expansion in the
inverse nucleon mass.
We then show results of detailed calculations and discuss
the implications of the presence of the nuclear potentials in the
interaction Hamiltonian.
In section \ref{FW} we discuss the FW transformation of a
relativistic Hamiltonian describing a particle interacting with
electromagnetic and strong nuclear fields, and we find the
corresponding nonrelativistic amplitude to second order in the
inverse nucleon mass.
We discuss differences between the interaction Hamiltonians
obtained through the Pauli and FW reduction schemes.
The section closes with a discussion of
observables calculated at different orders, with and without
potentials in the FW interaction Hamiltonian.
Section \ref{conclusion} is devoted to conclusions.

\section{Relativistic Direct Knock-out Mechanism}
\label{relativistic}

The relativistic distorted wave amplitude (S matrix) for the reaction
$\left(\gamma,p\right)$
on a target nucleus in the angular momentum state
$\left| J_{i} M_{i} \right> $ leading to a residual nuclear
state $\left| J_{f} M_{f} \right> $ is obtained in first order
in the interaction Hamiltonian as
\cite{LS88}
\begin{eqnarray}
  S_{fi} &=& \frac{ - i } { \left( 2 \pi \right)^{3} }
             \left[ \frac{1}{2 \omega} \right ]^{1/2}
             \left[ \frac{M}{E_{\scriptsize C}} \right ]^{1/2}
     \nonumber \\
      & & \times
             \sum_{ J_B M_B}{
               \left( J_f, J_B; M_f, M_B| J_i, M_i \right)
               \left[ {\cal S}_{J_i J_f} (J_B) \right]^{1/2}  }
     \nonumber \\
      & & \hspace{14 mm} \times
               \int{ \overline{\psi}^{\left( - \right)}
                                    _{{\scriptsize S}_{f}}
                     \left( x \right)
                     H_{em} \left( x \right)
                     \psi_{J_{B} M_{B}} \left( x \right)
                     d^{4} x }      ,
 \label{rel_s_mat}
\end{eqnarray}
where $E_{\scriptsize C}$ is the energy of the final state
continuum proton
and $M$ is its mass. The energy of the incident photon is
$\omega$.
The Clebsch-Gordan coefficient in equation
(\ref{rel_s_mat}) couples
the bound proton with angular momentum $J_B$ and projection
$M_B$ to the residual nucleus,
and ${\cal S}$ is the spectroscopic factor.
The electromagnetic interaction Hamiltonian is
\begin{eqnarray}
   H_{em} \left( x \right) =
           e { / \hspace{-0.115 in} A } \left( x \right)
         + \frac{\kappa}{2} \frac{e}{2M} \sigma^{\mu \nu}
           F_{\mu \nu} \left( x \right)   .
  \label{h_em}
\end{eqnarray}
Notice that in this work we do not consider modifications to
the electromagnetic interaction Hamiltonian arising when a
nucleon is off-shell \cite{NK87}.
The 4-vector potential describing the photon is written as
\begin{eqnarray}
    A^{\mu}_{\xi} \left( x \right)
%          = \frac{1}{\left( 2 \pi \right)^{3/2}}
           =\epsilon^{\mu}_{\xi} e^{-ik_{\gamma} \cdot x} ,
\label{em-pot}
\end{eqnarray}
where the label $\xi$ is the polarization angle of the photon,
allowing us to choose between two linear polarization states
when the polarization 4-vector is written as
$\epsilon^{\mu}_{\xi} = \left( 0, cos\xi, sin\xi, 0 \right)$.
In writing the polarization vector this way we have made a
definite choice of z-axis; namely, the z-axis is taken along
the direction of the photon momentum so the photon
4-momentum is always
$k^{\mu}_{\gamma} = \left( E_{\gamma}, 0, 0, k_{\gamma} \right)$.
We thus have two linear polarization states: $\xi = 0^{\circ}$
with polarization along the x-axis and $\xi = 90^{\circ}$ with
polarization along the y-axis.

The second term in the interaction Hamiltonian (\ref{h_em}) is
an anomalous magnetic moment term with $\kappa = 1.79$ for
the proton.
The tensor $\sigma^{\mu\nu}$ is related to the usual
Dirac gamma matrices through,
\begin{eqnarray}
   \sigma^{\mu \nu} = \frac{i}{2}
                      \left[ \gamma^\mu, ~ \gamma^\nu \right]  ,
\label{1.6}
\end{eqnarray}
and the electromagnetic field tensor $F_{\mu \nu}$ is
\begin{eqnarray}
   F_{\mu \nu} = \partial_\mu A_\nu - \partial_\nu A_\mu  .
 \label{1.7}
\end{eqnarray}

The Dirac spinors in equation (\ref{rel_s_mat}) describe the bound
and continuum nucleons and are solutions of a Dirac equation
of the form
\begin{eqnarray}
   \left\{   \mbox{\boldmath{$\alpha$}} \cdot
             \mbox{\boldmath{$p$}}
           + \beta \left[ M + S \left( r \right) \right]
           + V \left( r \right)
          \right\} \psi \left( x \right)
   = E \psi \left( x \right),
  \label{1.2}
\end{eqnarray}
where we adopt the standard representation of  the $4 \times 4$
Dirac matrices $\left\{ \alpha_{i} \right\}$
and $\beta$ \cite{BDQM}. The potentials
$S \left( r \right)$ and $V \left( r \right)$
are the scalar and zeroth-component vector potentials respectively.
These potentials are originally derived from the effective
Lagrangian of the $\sigma +\omega$ model \cite{SW}.
In actual distorted wave calculations, the final state proton is
described by a solution of the Dirac equation above,
containing complex potentials
$S_{\scriptsize C} \left( r \right)$ and
$V_{\scriptsize C} \left( r \right)$.
The parameters for these potentials are obtained through
analysis of data from proton elastic scattering on nuclei \cite{EC}.
For the bound nucleon, Dirac-Hartree potentials are used \cite{HS}.
Detailed discussions of the Dirac equations pertaining to both
the bound and continuum nucleons can be found in references
\cite{LS88, HS}.

In the rest of this section we discuss a Schr\"{o}dinger-like wave
equation derived from the Dirac equation (\ref{1.2}).
The solutions of this Schr\"{o}dinger-like
equation will be used in the next section to cast the amplitude
in a nonrelativistic form.

We write the Dirac spinors in terms of upper and
lower components $u$ and $\ell$
\begin{eqnarray}
   \psi \left( x \right)
        = \left[ \matrix{ u \left( x \right) \cr
                          \ell \left( x \right) \cr } \right]   ,
 \label{1.8}
\end{eqnarray}
and use the Dirac equation (\ref{1.2}) to write the lower
component of the wave function in terms of the upper component as
\begin{eqnarray}
   \ell \left( x \right)
     = \frac{\mbox{\boldmath{$\sigma$}} \cdot
             \mbox{\boldmath{$p$}}}
            {E + M + S \left( r \right) - V \left( r \right)}
       \, u \left( x \right)   .
  \label{1.9}
\end{eqnarray}
Thus the Dirac spinors can be written in terms of their upper
components in the form
\begin{eqnarray}
   \psi \left( x \right)
         = \left[   \matrix{   1 \cr
                    \mbox{\boldmath{$\sigma$}} \cdot
                    \mbox{\boldmath{$ p$}} \vspace{0 mm} \cr
                    \overline{M + E + S\left( r \right)
                                - V\left( r \right)}   \cr   }
                  \right] u \left( x \right) .
          \label{1.10}
\end{eqnarray}

The upper component of the Dirac spinor $u \left( x \right)$
can be related to a Schr\"{o}dinger-like wave function
$\Psi_{\mbox{{\scriptsize Sch}}} \left( x \right)$ by
\cite{CHM}
\begin{eqnarray}
   u \left( x \right) = D^{\frac{1}{2}} \left( r \right)
                        \Psi_{\mbox{{\scriptsize Sch}}}
                        \left( x \right)   ,
  \label{sch_trans}
\end{eqnarray}
and the function $D \left( r \right)$ depends on the
Dirac potentials as
\begin{eqnarray}
   D \left( r \right) =\frac{E + M + S \left( r \right)
                                    - V \left( r \right)}
                             {E + M}
                      =\frac{s\left( r \right)}{E + M}    ,
 \label{1.11}
\end{eqnarray}
Note that $D(r)$ goes to one for large $r$, so the asymptotic
behavior of the upper component of the
Dirac wave function and the Schr\"{o}dinger-like
wave function are the same.
The Schr\"{o}dinger-like wave function
$\Psi_{\mbox{{\scriptsize Sch}}} \left( x \right)$ is the
solution of the Schr\"{o}dinger-like equation \cite{CHM}
\begin{eqnarray}
    \left\{ - \frac{\mbox{\boldmath{$\nabla$}}^{2}}{2 M}
            + U_{\mbox{{\scriptsize cent}}} \left( r \right)
            + U_{\mbox{{\scriptsize so}}} \left( r \right)
              \mbox{\boldmath{$\sigma$}} \cdot \mbox{\boldmath{$L$}}
           \right\} \Psi_{\mbox{{\scriptsize Sch}}}
         = \left( E - M \right) \Psi_{\mbox{{\scriptsize Sch}}}   .
  \label{1.12}
\end{eqnarray}
The central and spin-orbit potentials are functions of the Dirac
potentials as well as the energy of the proton, and
are written explicitly as
\begin{eqnarray}
   U_{\mbox{{\scriptsize cent}}} \left( r \right)
      &=& E - M
          + \frac{1}{2 M}
            \left\{   s d
                    - \frac{s^{\prime}}{sr}
                    - \frac{1}{2} \frac{s^{\prime \prime}}{s}
                    + \frac{3}{4} \frac{{s^{\prime}}^{2}}{s^{2}}
                   \right\}   ,
  \label{1.14a}
\end{eqnarray}
and
\begin{eqnarray}
   U_{\mbox{{\scriptsize so}}} &=& - \frac{1}{2 M}
                          \frac{s^{\prime}\left( r \right)}
                           {rs\left( r \right)}   .
  \label{1.14b}
\end{eqnarray}
In addition to the function $s(r)$ defined in equation (\ref{1.11}),
we have defined a function involving the difference of the nucleon's
mass and  energy
\begin{eqnarray}
     d \left( r \right) = M - E + S \left( r \right)
                                + V \left( r \right)   .
  \label{1.15b}
\end{eqnarray}
For the bound state this ``nonrelativistic'' equation provides
a better description of spin orbit splitting than the usual
nonrelativistic calculations \cite{HS},
and similarly for the continuum nucleon the
wave function obtained from this equation gives an improved
description of nucleon-nucleus elastic scattering data \cite{EC}.

\section{Effective Pauli Reduction}
\label{Pauli}

Pauli reduction of the relativistic amplitude for the knock-out
contribution to $\left(\gamma,p\right)$ reactions was described briefly
in reference
\cite{HS94}; we provide more details and discussion here.
In the effective Pauli reduction scheme the relativistic distorted
wave S matrix is rewritten in terms of Schr\"{o}dinger-like wave
functions \cite{CHM} resulting in an effective interaction Hamiltonian
which may be expanded in powers of $\frac{1}{E+M}$.
The various orders can then be related, in the proper limit, to the
nonrelativistic form of the amplitude. As we show below there are,
however, important differences with the usual nonrelativistic amplitude.
The qualitative features of a similar two component reduction of the
Dirac wave function for several generic vertices, in the presence of the
nuclear interactions, has been discussed by Cooper {\it et al.}
\cite{COMAC83}.
The emphasize in our present discussion is to present a more quantitative
analysis for the case of $\left(\gamma,p\right)$ reactions.

\subsection{Formalism}
Beginning with the amplitude of equation (\ref{rel_s_mat}),
we write the integral in terms of Schr\"{o}dinger-like wave
functions using equations (\ref{1.10}) and (\ref{sch_trans})
relating the Dirac spinors to the Schr\"{o}dinger-like wave
functions. This allows us to write
\begin{eqnarray}
  S_{fi} &=& \frac{ - i } { \left( 2 \pi \right)^{3} }
             \left[ \frac{1}{2 \omega} \right ]^{1/2}
             \left[ \frac{M}{E_{\scriptsize C}} \right ]^{1/2}
     \nonumber \\
         & & \times
             \sum_{ J_B M_B}{
             \left( J_f, J_B; M_f, M_B| J_i, M_i \right)
             \left[ {\cal S}_{J_i J_f} (J_B) \right]^{1/2} }
     \nonumber \\
         & & \hspace{14 mm} \times
        \int{\Psi^{\dagger\left( - \right)}_{{\mbox{{\scriptsize Sch,S}}}_{f}}
             \left( x \right)
              H^{eff}_{em} \left( x \right)
             \Psi_{{\mbox{{\scriptsize Sch}}},J_{B} M_{B}}
             \left( x \right)
              d^{4} x }      ,
 \label{nonrel_s_mat}
\end{eqnarray}
where the effective interaction Hamiltonian $H^{eff}_{I}$ is
\begin{eqnarray}
 H^{eff}_{I} &=& \matrix{ \cr \cr \cr }
                   D_{\scriptsize C}^{1/2}\left( r \right)
                   \left[  1\; \;
                   \frac{ \mbox{\boldmath{$\sigma$}} \cdot
                   \mbox{\boldmath{$p$}} }
                   {M + E_{\scriptsize C} + S_{\scriptsize C}
                   \left( r \right)
                    - V_{\scriptsize C}\left( r \right)}
                                \right]\gamma^{0} H_{em}(x)
       \nonumber \\
              & & \hspace{8.5 mm} \times \;\;\;\;\;
                   \left[   \matrix{   1   \cr
                  \mbox{\boldmath{$\sigma$}} \cdot
                  \mbox{\boldmath{$ p$}}  \vspace{0 mm}    \cr
                  \overline{M + E_{\scriptsize B}
                  + S_{\scriptsize B}\left( r \right)
                  - V_{\scriptsize B}\left( r \right)}
                  \cr   } \right]
                  D_{\scriptsize B}^{1/2}\left( r \right)  .
\label{h_eff}
\end{eqnarray}
This can be expanded in powers
of $\frac{1}{E+M}$ and written in the form:
\begin{equation}
 H^{eff}_I= H_I^{(1)}+ H_I^{(2)}+\cdots  .
\label{1.18}
\end{equation}
The first and second order contributions are given by
\begin{eqnarray}
  H_I^{(1)} = &-& \frac{e\kappa }{2M}
                       \mbox{\boldmath{$\sigma$} }
                       \cdot \mbox{\boldmath{$ \nabla$}}\times
                       \left[{\bf A} \right]
                       -e\left(\frac{1}{M+E_{\scriptsize C}}
                       +\frac{1}{M+E_{\scriptsize B}}
                        \right){\bf A}\cdot
                       {\bf p}
  \nonumber\\ \nonumber
                   &+& ie\left(\frac{1}{M+E_{\scriptsize C}}
                       -\frac{1}{M+E_{\scriptsize B}}\right)
                       \mbox{\boldmath{$\sigma$}}
                       \cdot {\bf A}\times{\bf p}
                       -e\frac{\mbox{\boldmath{$\sigma$}}
                       \hspace{-.05in}\cdot\hspace{.05in}
                       \mbox{\boldmath{$ \nabla$}} \times
                       \left[ {\bf A} \right]}{M+E_{\scriptsize C}}
  \\ \nonumber \\ \nonumber
   H_I^{(2)} = &-& \frac{e}{2}\left[\frac{\kappa}{2M}
                       \mbox{\boldmath{$\sigma$} }\hspace{-.05in}
                       \cdot\hspace{.05in}\mbox{\boldmath{$ \nabla$}}
                       \times \left[ {\bf A} \right]
                      +\frac{\mbox{\boldmath{$\sigma$}}
                       \hspace{-.05in}\cdot\hspace{.05in} {\bf A}
                       \mbox{\boldmath{$\sigma$} }\hspace{-.05in}
                       \cdot\hspace{.05in}{\bf p}}
                       {M+E_{\scriptsize B}}\right]
                       Q_{\scriptsize B}(r)
  \\ \nonumber
                   &+& e\left[Q_{\scriptsize B}(r)
                       -\kappa\frac{1}{2M}\omega\right]
                       \mbox{\boldmath{$\sigma$} }\hspace{-.05in}
                       \cdot\hspace{.05in} {\bf A}
                       \frac{\mbox{\boldmath{$\sigma$} }\hspace{-.05in}
                       \cdot\hspace{.05in}{\bf p}}{M+E_{\scriptsize B}}
  \\ \nonumber
                   &-& \frac{e}{2}Q_{\scriptsize C}(r)
                       \left[\frac{}{}\frac{\kappa}{2M}
                       \mbox{\boldmath{$\sigma$} }\hspace{-.05in}
                       \cdot\hspace{.05in}\mbox{\boldmath{$ \nabla$}}
                       \times \left[ {\bf A}
                       \right]+\frac{\mbox{\boldmath{$\sigma$}}
                       \hspace{-.05in}\cdot\hspace{.05in} {\bf A}
                       \mbox{\boldmath{$\sigma$} }\hspace{-.05in}
                       \cdot\hspace{.05in}{\bf p}}
                       {M+E_{\scriptsize B}}\right.
 \\ \nonumber
                  & & \left.\hspace{1.8 in}+\frac{\mbox{
                      \boldmath{$\sigma$} }\hspace{-.05in}\cdot
                      \hspace{.05in}{\bf p}\mbox{\boldmath{$\sigma$}}
                      \hspace{-.05in}\cdot\hspace{.05in}
                      {\bf A}}{M+E_{\scriptsize C}}\frac{}{}\right]
 \\
                  &+& e\frac{\mbox{\boldmath{$\sigma$}}
                      \hspace{-.05in}\cdot\hspace{.05in}
                      {\bf p}}{M+E_{\scriptsize C}}
                      \left\{Q_{\scriptsize C}(r)
                     -\frac{1}{2}Q_{\scriptsize B}(r)
                     +\frac{\kappa}{2M}\omega\right\}
                      \mbox{\boldmath{$\sigma$} }\hspace{-.05in}
                      \cdot\hspace{.05in} {\bf A}  ,
\label{1.19}
\end{eqnarray}
where we have written
\begin{eqnarray}
     Q_{\scriptsize X}(r)
        = \frac{S_{\scriptsize X}(r) - V_{\scriptsize X}(r)}
               {E+M}  .
\end{eqnarray}
${\bf O}\left[ f \right]\cdots$  in the interaction Hamiltonians
of equation (\ref{1.19}) means that operator ${\bf O}$
acts only on function $f$.
The appearance of the Dirac potentials through Q's in the
interaction Hamiltonians (\ref{1.19}) delineates the modification
of the effective photon interaction due to the presence of the
nuclear medium.
We will investigate the significance of this medium effect in the
following section.

Using the Hamiltonian (\ref{1.19}) in equation (\ref{nonrel_s_mat})
along with the nonrelativistic wave functions, we can cast the amplitude in
a nonrelativistic form as follows
\begin{eqnarray}
  S^{(i)}_{fi} &=& \frac{-i}{(2\pi)^3}
             \left[ \frac{1}{2 \omega} \right]^{1/2}
             \sum_{ J_B M_B}{
             \left( J_f, J_B; M_f, M_B| J_i, M_i \right)
             \left[ {\cal S}_{J_i J_f} (J_B) \right]^{1/2} }
     \nonumber \\
         & & \hspace{14 mm} \times
        \int{\Psi^{\dagger\left( - \right)}_{{\mbox{{\scriptsize Sch,S}}}_{f}}
             \left( x \right)
              H^{(i)}_{I} \left( x \right)
             \Psi_{\mbox{{\scriptsize Sch}},J_{B} M_{B}} \left( x \right)
              d^{4} x }      ,
 \label{non1-rel_s_mat}
\end{eqnarray}
where $(i)$ refers to the highest order of the inverse of the nucleon mass
in the interaction Hamiltonian used in the nonrelativistic amplitude.
The amplitude (\ref{non1-rel_s_mat}), with $ H_I^{i}= H_I^{(1)}$ in the
limit $E_{\scriptsize B},E_{\scriptsize C}\rightarrow M$ is equivalent
to the usual nonrelativistic transition amplitude \cite{BGP81} except
that the Schr\"{o}dinger-like wave functions are used instead of the usual
nonrelativistic Schr\"{o}dinger wave functions.
The amplitude obtained this way will be referred to here as the
first order nonrelativistic amplitude.

The Schr\"{o}dinger-like wave functions describing the bound nucleon
can be written as
\begin{eqnarray}
\Psi_{\mbox{{\scriptsize Sch,B}}}(x) &=& e^{-iE_{B}t}f_{L_{B}}(r)
                                              {\cal Y}^{M_{B}}
                                              _{L_{B}1/2J_{B}}(\Omega) ,
 \label{1.20-1}
\end{eqnarray}
while for the continuum nucleon we write
\begin{eqnarray}
\Psi_{\mbox{{\scriptsize Sch,C}}}^{\dagger}(x)
                             &=&4\pi e^{iE_{\scriptsize C}t}
                                 \sum_{LMJ}i^{-L}
                                 Y^{M-S_{f}}_{L}(\hat{k}_{f})
 \nonumber\\
                       & &\hspace{.5 in}\times
                      (L, 1/2; M-S_{f}, S_{f}|J, M)
                      f_{LJ}(r){\cal Y}^{M\dagger}_{L1/2J}(\Omega)  ,
 \label{1.20-2}
\end{eqnarray}
where
\begin{eqnarray}
{\cal Y}^{M}_{L1/2J}(\Omega)
                            &=& \sum_{\mu}(L, 1/2; M-\mu, \mu |J, M )
                                Y^{ M-\mu}_{L}(\Omega )
                                \chi ^{\mu }_{1/2} .
\label{1.20-a}
\end{eqnarray}
It should be emphasized here that the wave functions introduced in
equations (\ref{1.20-1}) and (\ref{1.20-2})
have nonrelativistic normalization, i.e the factor
$\sqrt{\frac{E+M}{2M}}$ which comes from the Dirac spinor describing
the outgoing nucleon has been set equal to one (thus a bound state
wave function is normalized to one and the
plane wave limit of the nucleon wave function is of the form $
\Psi_{\mbox{{\scriptsize Sch,S}}_f}(x)=e^{-ik\cdot x}
                            \chi ^{\mbox{{\scriptsize S}}_f}_{1/2}$).
In addition the factor $\sqrt{\frac{M}{E}}$ which comes from the
Dirac field expansion is set equal to one in the cross section.

Using equations (\ref{em-pot}), (\ref{1.20-1}) and (\ref{1.20-2}) then
after evaluating the angular integration the first order amplitude
can be written as
\begin{eqnarray}
\hspace{-.3 in}   S_{fi}^{(1)} &=& \frac{-ie}{\pi}
                   \left[ \frac{1}{2 \omega} \right ]^{1/2}
                   \sum_{ J_B M_B}{
                   \left( J_f, J_B; M_f, M_B| J_i, M_i \right)
                   \left[ {\cal S}_{J_i J_f} (J_B) \right]^{1/2} }
  \nonumber\\ \nonumber
              & & \hspace{.01 in}\times\delta(E_{\scriptsize C}
                  -E_{\scriptsize B}-\omega)
                  \sum_{lLJ\mu}(-i)^{l+L}(2l+1)
  \\ \nonumber
              & & \hspace{.01 in}\times
                  \frac{{}}{{}}\left\{\frac{{}}{{}}
                  (L,1/2;M_B+2\mu-S_f,S_f
                  \mid J,M_B+2\mu)
                   Y^{M_B+2\mu-S_f}_L(\hat{k}_f)
                  \right.
  \\ \nonumber
              & & \hspace{.15 in}\times
                  \left[ \left(\frac{\kappa\omega}{2M}
                  +\frac{\omega}{M+E_{\scriptsize C}}\right)
                   I_{l,L,J,L_B}C^{\mu}_{l,L,J,L_B}\right.
  \\ \nonumber
              & &  \hspace{.4in}-(2\mu\cos{\xi}-i\sin{\xi})
	           \left(\frac{1}{M+E_{\scriptsize C}}
                  -\frac{1}{M+E_{\scriptsize B}}\right)
  \\ \nonumber
              & & \hspace{1.8in}\left.
                  \times{\cal H}^{M_B,\mu}_{LJL_B}
                  (P^{M_B,\mu,0}_{L_B+1,l,L}
                 -{\cal P}^{M_B,\mu,0}_{L_B-1,l,L})
                  \frac{}{}\right]
  \\ \nonumber
              & & \hspace{.25 in}+\sum_{\nu=\pm 1}
                  (L,1/2;M_B-\nu-S_f,S_f
                  \mid J,M_B-\nu)Y^{M_B-\nu-S_f}_L(\hat{k}_f)
  \\ \nonumber
              & & \hspace{.4in}\times\left[\left(\frac{1}
                  {M+E_{\scriptsize C}}
                 +\frac{1}{M+E_{\scriptsize B}}\right)
                 +2\nu\mu\left(\frac{1}{M+E_{\scriptsize C}}
                 -\frac{1}{M+E_{\scriptsize B}}\right)\right]
  \nonumber\\
              & & \hspace{.4in}\left.\times H^{M_B,\mu,\nu}_{L,J,L_B}
                  [P^{M_B,-\mu,\nu}_{L_B+1,l,L}-{\cal P}^{M_B,-\mu,\nu}
                  _{L_B-1,l,L}]\frac{\nu\cos{\xi}+i\sin{\xi}}{\sqrt{2}}
                  \frac{}{}
                  \right\}  ,
\label{1.23}
\end{eqnarray}
where $ I_{l,L,J,L_B}$, $C^{\mu}_{l,L,J,L_B}$,
$P^{M_B,\mu,\nu}_{L_B+1,l,L}$,
$ {\cal P}^{M_B,\mu,\nu}_{L_B-1,l,L}$,
$H^{M_B\mu,\nu}_{L,J,L_B}$,
and ${\cal H}^{M_B,\mu}_{L,J,L_B}$ involve radial integrals and
Clebsch-Gordan coefficients.
These functions are defined in appendix A.
The corresponding expression for the amplitude to second order
in $\frac{1}{E+M}$ has the same
structure as (\ref{1.23}) but the contributing terms are more
complicated.

\subsection{Results of the Effective Pauli Reduction}
\label{Pauli-res}

In the previous subsection we illustrated how the effective Pauli
reduction of the relativistic amplitude for the knock-out contribution
to $(\gamma,p)$ reactions is performed to get an expansion in powers of
$\frac{1}{E+M}$.
The successive terms in this expansion can be reduced,
in the appropriate limits,
to forms that are equivalent to the amplitudes used in nonrelativistic
calculations.
This allows us to carry out quantitative comparisons between the
relativistic and nonrelativistic calculations.
The appropriate limits referred to above include
i) setting the nucleon total energy equal to its rest mass
ii) turning off the nuclear potentials in the second-order
interaction Hamiltonian
iii) taking proper account of wave function normalizations.
This comparison will be carried out for the differential cross section
as well as the photon asymmetry \cite{GGZKRS} at representative
energies.
We shall compare the following four types of calculations:
\begin{quote}
a) Full relativistic calculations using the amplitude
given by equation (\ref{rel_s_mat}).
In the figures that follow these calculations are represented by
solid curves and denoted ``Relativistic''.
For these calculations the relativistic $(\gamma,p)$
code of Lotz has been used \cite{GE}.
\\
b) First order nonrelativistic calculations.
These calculations are obtained using the amplitude
$S^{(1)}_{fi}$ of equation (\ref{1.23}).
Note that in the limit $E_{\scriptsize B}$ and $E_{\scriptsize C}
\rightarrow M$,
the interaction Hamiltonian takes on a simplified form.
These calculations essentially represent the standard nonrelativistic
calculations. Comparison of these results with the relativistic
calculations gives the essence of the difference between the two approaches.
These calculations are denoted ``Pauli N.R. (First order)'' in Figs. 1 and 2
and are represented by the dotted curves.
\\
c) The third type of calculation represents a nonrelativistic
calculation carried out to second order in the inverse nucleon mass.
This calculation then includes the interaction Hamiltonian
$H^{(2)}_I$, but with the nuclear vector and scalar potential set
equal to zero in the interaction Hamiltonian.
This interaction will then relate to the limit in which the photon is
interacting with a free nucleon. Our intention here is to clarify how
much improvement in the nonrelativistic calculations can be obtained
by including second order effects.
We shall show that the effects are not substantial. These results are
shown by the dot-dashed curves in Figs. 1 and 2 and are
labelled ``Pauli N.R. (First + Second)''. We shall refer to these
calculations in the text as ``{\it medium-uncorrected}'' calculations
to signify the fact that they pertain to the limit in which the
nuclear potentials are set to zero.
\\
d) The fourth type of calculation represents a nonrelativistic
calculation using the full expression for $H^{(2)}_I$, i.e.
with the effect of the nuclear medium (through the presence of
the potentials) taken into account.
These calculations are shown by the dashed curves in Figs. 1 and 2
and are labelled ``Pauli N.R. (First + Full Second)''.
They will be referred to as ``{\it medium-corrected}''
nonrelativistic calculations in the following text .
The essence of the present comparison is to show the significance
of these medium effects.
\end{quote}

Since our aim in this paper is to compare the two
theoretical models, namely relativistic and nonrelativistic,
for the knock-out contribution to $\left( \gamma, p \right)$ reactions,
we do not compare the resulting observables with data.
We refer the reader to the work of Lotz and Sherif \cite{LS88}
for comparison of the results of the relativistic model with data.

The bound state wave functions used in the calculations
reported in this section are generated using the Dirac-Hartree
potentials of Horowitz and Serot \cite{HS}.
In all the calculated cross sections given in this paper
the spectroscopic factor takes its maximum allowable value of
($2J_B+1$) for
both the relativistic and nonrelativistic calculations.

Figure 1 shows the calculated observables for the
$^{16}O(\gamma,p)~^{15}N$ reaction
with a photon energy of $E_\gamma =100 $ MeV. The residual nucleus
is in a $1p_{\frac{1}{2}}$ single hole state. The final
state optical potential
is taken from reference  \cite{GE}.
The cross section curves of Fig. 1(a) show large
differences between the first order nonrelativistic
(dotted~ curve) and the relativistic
calculations (solid~curve). At forward angles
the nonrelativistic
calculations are almost an order of magnitude larger
than the relativistic calculations, while at backward
angles the nonrelativistic calculations are roughly two orders of
magnitude smaller than the relativistic calculations.
Medium-uncorrected second order calculations improve
the nonrelativistic calculations slightly only at forward
angles (dot-dashed~curve).
On the other hand the medium-corrected second order calculations
produce large changes in the nonrelativistic calculations.
Note the large change in the magnitude for both small and
large scattering angles (dashed~curve) which brings
the nonrelativistic calculations into close agreement
with the results of the relativistic model.

Calculations of photon asymmetry shown in Fig. 1(b) also
exhibit noticeable differences between the first order
nonrelativistic and relativistic calculations.
These two calculations have different shapes and magnitudes
especially for scattering angles greater than 40$^\circ$.
Medium-uncorrected second order calculations modify the nonrelativistic
calculations slightly in magnitude while the medium-corrected second order
calculations result in noticeable changes in the shape and
magnitude of the nonrelativistic calculations.
These changes are such that at forward angles the nonrelativistic
calculations now overlap the relativistic calculations and are much closer
in shape and magnitude at backward angles.

Figure 2 shows similar comparisons for the
reaction at a higher incident photon energy, $E_\gamma =312$ MeV.
The final state global optical potentials are taken from
Cooper {\it et al.} \cite{EC}. Figure 2(a) shows that
the cross section obtained from the first order nonrelativistic
calculations has both different shape
and magnitude from the results of the relativistic
calculations. The nonrelativistic calculations
lie above the relativistic calculations for angles
smaller than 40$^\circ$, whereas for other scattering angles
the nonrelativistic calculations lie below the relativistic
calculations by as much as an order of magnitude.
Medium-uncorrected second order calculations
lie closer to the relativistic
calculations only at forward angles.
Medium-corrected second order calculations,
on the other hand, are much closer to the results of the
relativistic calculations.

Calculations of the
photon asymmetries of Fig. 2(b) show that the first order
nonrelativistic results differ from the relativistic
calculations in both shape and magnitude.
Medium-uncorrected second order calculations improve the nonrelativistic
calculations slightly at forward angles but the overall shape
stays the same as that of the first order.
Medium-corrected second order calculations modify the shape and
magnitude of the nonrelativistic calculations at all scattering angles.
It might not be clear visually that these changes
bring the nonrelativistic calculations into noticeably better agreement
with the
relativistic ones, however a chi-squared comparison does indeed show that the
medium-corrected second order calculations are
closer to the relativistic calculations than medium-uncorrected second
order calculations.
The important point here is that there are large differences
in the calculations when medium corrections are taken into account.

{}From these examples one can see that the first order nonrelativistic
calculations are different from relativistic calculations.
Attempts to improve the situation through the inclusion of the second
order terms in which the presence of the nuclear potentials is neglected
(as would normally be done in typical nonrelativistic calculations)
are bound to fail in bringing the results close to the relativistic
calculations.
We have seen that medium-corrected second order calculations
(i.e. those that include the effects of the nuclear potentials
on the interaction Hamiltonian) are much closer.
We remind the reader that these potentials are absent in interaction
terms of the usual nonrelativistic calculation even when higher order
relativistic corrections are included \cite{BGP81}.
This medium modification is the important ingredient that is missing in
ordinary nonrelativistic calculations.
It is this medium modification that is responsible for many of the
differences between the two types of calculations.
In the following section we show that the same conclusion can be
reached through a procedure based on the Foldy Wouthuysen transformation.

\section{Foldy-Wouthuysen Transformation}
\label{FW}

Following McVoy and Van Hove \cite{MVVH} many authors construct a
nonrelativistic model of photons interacting with
 nuclei by performing a FW transformation on the
relativistic Hamiltonian which involves the electromagnetic
interaction with a {\it free} nucleon. The resulting nonrelativistic
interaction Hamiltonian is then sandwiched between
Schr\"{o}dinger wave functions describing the initial and final
nucleons \cite{BGP81}. To investigate the effect of the nuclear medium
on the FW reduction of the relativistic amplitude of the knock-out
contribution to $(\gamma,p)$ reactions, we perform a FW
transformation on the relativistic Hamiltonian of a
photon interacting with a nucleon in the presence of strong scalar and
vector potentials.

In the preceding discussion of the Pauli reduction scheme,
our starting point was the distorted wave amplitude of equation
(\ref{rel_s_mat}).
A feature of this amplitude for practical calculation is the
use of complex vector and scalar potentials to describe the
interaction of the outgoing nucleon with the residual nucleus.
The unitary requirement for the FW transformation makes it
unacceptable to work with this type of amplitude
(note that in the usual DWBA amplitude the initial and
final states of the nucleon are described by different Hamiltonians).
We must therefore work with a model amplitude in which the
bound and continuum state potentials are the same and real.
It is known that such model amplitudes are inferior in their
description of the data in comparison to the distorted wave amplitudes.
It must be noted however that the purpose of the present
investigation is not aimed at fitting data;
rather we are interested in features that differentiate between
the relativistic and nonrelativistic calculations.
For this purpose the restricted model amplitude
used here is quite appropriate.

\subsection{Spin $\frac{1}{2}$ Particle Interacting with Strong
and Electromagnetic Potentials }

We first summarize the FW transformation for the
case in which a Dirac particle interacts with a general field
following the procedure as given in the reference \cite{BDQM}.
The results will then be applied to the case in which a
nucleon interacts with an electromagnetic field while under the
influence of the strong potentials.
The Dirac equation is written in the general form
\begin{eqnarray}
    i \partial_{t} \, \psi \left( x \right) = H \psi \left( x \right)  ,
\label{1-2-3}
\end{eqnarray}
where the relativistic Hamiltonian $H$ can be written in
 terms of even and odd operators as
\begin{eqnarray}
     H = \beta \, M + {\cal E} + {\cal O}  .
\label{1-2-4}
\end{eqnarray}
The odd operator ${\cal O}$ connects the upper component of the
Dirac spinor to the lower component while the even operator ${\cal E}$
can only connect either upper or lower components.
Now we perform the FW transformation on the relativistic
Hamiltonian (\ref{1-2-4}) following reference \cite{BDQM}.
After three successive transformations we find.
\begin{eqnarray}
          H^{\prime} &=& \beta \, M + {\cal E}
                         + \frac{\beta}{2M} {\cal O}^{2}
                         - \frac{i}{8M^{2}} [ {\cal O}, \dot{\cal O} ]
 \nonumber \\
                     & & \hspace{8 mm}
                         - \frac{1}{8M^{2}} \left[ {\cal O},
                         \left[ {\cal O}, {\cal E} \right] \right]
                         - \frac{\beta}{8M^3} {\cal O}^{4}  .
\label{1-2-21}
\end{eqnarray}
(For simplicity the transformed Hamiltonian is denoted
$H^{\prime}$, this is in fact the same as the Hamiltonian
$H^{\prime\prime\prime}$ in reference \cite{BDQM}.)
This result applies to any pair of even and odd operators.
We are interested here in a transformed Hamiltonian containing terms
which are even up to second order in the inverse of the nucleon mass
so we will eventually drop the last term in equation (\ref{1-2-21}).

The relativistic Hamiltonian for a proton interacting
with an external electromagnetic field while under the influence of the strong
vector  and scalar nuclear potentials can be written as
\begin{eqnarray}
                H = \mbox{\boldmath{$\alpha$}} \cdot
                    \mbox{\boldmath{$\pi$}}
                    + e \Phi
                    + \beta \left[ M + S \! \left( r \right) \right]
                    + V \! \left( r \right)
                    + \frac{e\kappa }{2M}
                    \beta \sigma^{\mu \nu} F_{\mu \nu}   ,
\label{1-2-22}
\end{eqnarray}
where $\mbox{\boldmath{$\pi$}}={\bf p}-e{\bf A}$.
The above Hamiltonian can be written in terms of the
electric and magnetic fields ${\bf E}~\mbox{and}~{\bf B}$ as
\begin{eqnarray}
               H = \mbox{\boldmath{$\alpha$}}
                   \cdot \mbox{\boldmath{$\pi$}}
                   + \beta \left[ M + S \! \left( r \right) \right]
                   + V \! \left( r \right)
                   + \frac{i\kappa e}{2M}
                     \left( \mbox{\boldmath{$\gamma$} }
                     \hspace{-.05in}\cdot\hspace{.05in}{\bf E}
                   + i \mbox{\boldmath{$\Sigma$} }
                     \hspace{-.05in}\cdot
                     \hspace{.05in}{\bf B} \right)  ,
\label{1-2-23}
\end{eqnarray}
where we have defined
\begin{eqnarray}
\mbox{\boldmath{$\Sigma$} }=\beta\mbox{\boldmath{$\sigma$} }  .
\end{eqnarray}
We rewrite the Hamiltonian (\ref{1-2-23}) in terms of even and odd
operators as in equation (\ref{1-2-4}).
In this case the even and odd operators are
\begin{eqnarray}
         {\cal E} &=& \beta S \! \left( r \right) + V \! \left( r \right)
                      - \frac{\kappa e}{2M}
                      \mbox{\boldmath{$\Sigma$} }\hspace{-.05in}
                      \cdot\hspace{.05in}{\bf B}  ,
\label{1-2-24}
\end{eqnarray}
and
\begin{eqnarray}
        {\cal O} &=& \mbox{\boldmath{$\alpha$}}
                     \cdot \mbox{\boldmath{$\pi$}}
                     + \frac{ie \kappa}{2 M} \mbox{\boldmath{$\gamma$}}
                     \cdot \mbox{\boldmath{$E$}}   .
\label{1-2-25}
\end{eqnarray}
These even and odd operators are
used in equation (\ref{1-2-21})
in order to get the FW transformed Hamiltonian for a nucleon
interacting with electromagnetic and strong potentials.
We remind the reader that the quality of the FW transformation
depends on the assumption that the potential depths are small
compared with the nucleon mass; $V,\,S \ll M$.
This is not the case for the Dirac potentials, and so
we are by no means guaranteed that the FW transformation
will yield a convergent series.
We will return to this point below.

The resulting Hamiltonian to second order in the inverse of the
nucleon mass is a $4 \times 4$  block diagonal matrix
\begin{eqnarray}
\hspace{1.5 in}H'=\left [\matrix{
H'_{11} & 0\cr
0 & H'_{22} \cr
}\right ] .
\label{1-2-27}
\end{eqnarray}
The upper left element of this matrix (i.e $H'_{11}$)
corresponds to the transformed Hamiltonian for positive energy
solutions of the Dirac wave function.
For the nonrelativistic limit we will use this part of the
Hamiltonian which we write as
\begin{eqnarray}
H'_{11}=H_0+ H_I ,
\label{1-2-27_1}
\end{eqnarray}
where $H_0$ involves strong potentials whereas $ H_I$
carries both strong and electromagnetic interactions: we will
treat the latter as a perturbation on the former.
These Hamiltonians are evaluated explicitly using the Coulomb
gauge.
$H_0$ can be written to second order in $1/M$ as
\begin{eqnarray}
              H_0  &=&  \frac{{\bf p}^2}{2M}+M+S(r)+V(r)
    \nonumber \\
                   & &- \frac{1}{4M^2r}
                        \left[ S'(r)-V'(r) \right]
                        \mbox{\boldmath{$\sigma$} }
                        \cdot{\bf L}
    \nonumber \\
                   & &  -\frac{1}{2M^2}\left\{S(r){\bf p}^2
                        +{\bf p}\left[ S(r) \right]\cdot{\bf p}
                        +\frac{{\bf p}^2
                        \left[ V(r)+S(r) \right]}{4}\right\}  .
\label{1-2-28}
\end{eqnarray}
The interaction Hamiltonian is written in orders of $1/M$ as
\begin{eqnarray}
      H_I  &=&   H_I^{(1)}+ H_I^{(2)}+\cdots  ,
\end{eqnarray}
where for the first and second orders we have
\begin{eqnarray}
  H_I^{(1)}&=&  -\frac{e}{M}{\bf A}\cdot{\bf p}
                       -\frac{e}{2M}(1+\kappa)
                       \mbox{\boldmath{$\sigma$} }
                       \hspace{-.05in}\cdot\hspace{.05in}
                       (\hspace{.05in}\mbox{\boldmath{$\nabla$} }
                       \times{\bf A })  ,
 \nonumber \\ \nonumber \\
   H_I^{(2)}&=&  \frac{e\omega}{8M^2}(1+2\kappa)
                       \left\{\frac{}{}  \mbox{\boldmath{$\sigma$} }
                       \hspace{-.05in}\cdot
                       \hspace{.05in}\mbox{\boldmath{$\nabla$} }
                       \times\left[ \mbox{\boldmath{$A$} } \right]
                       -2i\mbox{\boldmath{$\sigma$} }
                       \hspace{-.05in}\cdot\hspace{.05in}{\bf A}
                       \times{\bf p}\frac{}{}\right\}
 \nonumber \\
                  & &  +\frac{e}{4M^2}\left\{\frac{}{} 2S(r){\bf A}
                       \cdot {\bf p}+2 {\bf A}\cdot {\bf p}S(r)
                       +2 S(r)\mbox{\boldmath{$\sigma$} }\hspace{-.05in}
                       \cdot\hspace{.05in}\mbox{\boldmath{$\nabla$} }
                       \times\left[ {\bf A } \right]\right.
 \nonumber \\
                  & &  \hspace{.75 in}+ \mbox{\boldmath{$\sigma$} }
                       \hspace{-.05in}\cdot\hspace{.05in}{\bf A}
                       \times\hspace{.05in}\mbox{\boldmath{$\nabla$} }
                       \left[ V(r)-S(r) \right]
 \nonumber\\
                  & &  \hspace{.75 in}\left.-
                       \left(\frac{}{} V(r)-S(r)
                       \frac{}{}\right)
                       \mbox{\boldmath{$\sigma$} }
                       \hspace{-.05in}\cdot\hspace{.05in}
                       \mbox{\boldmath{$\nabla$} }
                       \times\left[ {\bf A} \right]
                       \frac{}{}\right\}  .
\label{1-2-29}
\end{eqnarray}
Note that in equation (\ref{1-2-29}) terms of order $e^{2}$ are
dropped, and as before ${\bf O}\left[ f \right]\cdots$ means that
operator ${\bf O}$ operates only on function $f$.

\subsection{Schr\"{o}dinger-like Wave Functions}
The wave functions describing either the bound or continuum nucleons
are obtained by solving the equation
\begin{eqnarray}
  H_0\Psi({\bf r}) = E \Psi({\bf r}),
\label{1-2-28a}
\end{eqnarray}
where $H_0$ is the Hamiltonian (\ref{1-2-28}) containing
 terms to second order in $1/M$, and E is the total
 energy of the nucleon. Note that $H_0$ contains a first derivative
term which can be eliminated using the transformation
\begin{eqnarray}
  \Psi({\bf r})=D^{-\frac{1}{2}}_{\mbox{\scriptsize FW}}(r)
                \Psi_{\mbox{{\scriptsize Sch}}}({\bf r})  ,
  \label{1-2-31}
\end{eqnarray}
where $D_{\mbox{\scriptsize FW}}=1-\frac{S(r)}{M}$. With this
choice of $D_{\mbox{\scriptsize FW}}$,
 the two functions $\Psi({\bf r})$ and
$\Psi_{{\mbox{{\scriptsize Sch}}}}({\bf r})$
have the same
 asymptotic form. After this transformation the wave equation
(\ref{1-2-28a}) takes the form of the Schr\"{o}dinger-like equation
(\ref{1.12}). The central and
spin-orbit potentials in the present case are
\begin{eqnarray}
        U_{\mbox{{\scriptsize cent}}}(r)=& &   \frac{1}
                           {D_{\mbox{\scriptsize FW}}}\left\{S(r)+V(r)
                          +\frac{1}{4M^2r}\left[S'(r)+V'(r)\right]\right.
 \nonumber \\
                    & &   \hspace{1. in}+\left.\frac{1}{8M^2}
                           \left[S''(r)+V''(r)\right]\right\}
 \nonumber \\
                    & &   +(E-M)\left\{1-\frac{1}
                           {D_{\mbox{\scriptsize FW}}(r)}\right\}
 \nonumber \\
                    & &   +\frac{1}{2Mr} \frac{D_{
                           \mbox{\scriptsize FW}}'(r)}
                           {D_{\mbox{\scriptsize FW}}(r)}
                          +\frac{1}{4M} \frac{D_{\mbox{
                           \scriptsize FW}}''(r)}
                          {D_{\mbox{\scriptsize FW}}(r)}
                          -\frac{1}{8M} \left(
                          \frac{D_{\mbox{\scriptsize FW}}'(r)}
                          {D_{\mbox{\scriptsize FW}}(r)}\right)^2  ,
 \nonumber \\ \nonumber \\
     U_{\mbox{{\scriptsize so}}}(r)=  & & \frac{1}
                                       {D_{\mbox{\scriptsize FW}}(r)}
                                       \left\{-\frac{1}{4M^2r}
                                       \left[S'(r)-V'(r)\right]\right\}  .
\label{1-2-34}
\end{eqnarray}
The nonrelativistic amplitude for the knock-out contribution to the
$(\gamma,p)$ reaction at the desired order of $1/M$
takes the same form as given in equation (\ref{non1-rel_s_mat}).
Note however that the second order amplitude obtained through the FW scheme
involves wave functions which are solutions of equation (\ref{1-2-28a}).
As we have seen above, this equation does not have the form of the
usual Schr\"{o}dinger wave equation because it contains a first
order derivative of the wave function.
To be consistent with the usual nonrelativistic formalism we rewrite
the second
order amplitude in terms of the Schr\"{o}dinger-like wave functions
$\Psi_{\mbox{{\scriptsize Sch}}}({\bf r})$ introduced in equation
(\ref{1-2-31}).
This requires that the interaction
Hamiltonian be modified to
\begin{eqnarray}
              H_I'=D_{\mbox{\scriptsize FW}}^ {-\frac{1}{2}}
                   H_ID_{\mbox{\scriptsize FW}}^{-\frac{1}{2}}  .
\label{1-2-34a}
\end{eqnarray}
With this modification the first order terms in the interaction
Hamiltonian will remain the same as $ H^{(1)}_I$ in (\ref{1-2-29}),
while the second order terms in the interaction
Hamiltonian become
\begin{eqnarray}
   H_I^{(2)'} &=&   \frac{e\omega}{8M^2}(1+2\kappa)
                         \left\{  \mbox{\boldmath{$\sigma$} }
                         \hspace{-.05in}\cdot\hspace{.05in}
                         \mbox{\boldmath{$\nabla$} }\times
                         \left[ \mbox{\boldmath{$A$} } \right]
                         -2\mbox{\boldmath{$\sigma$} }
                         \hspace{-.05in}\cdot\hspace{.05in}{\bf A}
                         \times\mbox{\boldmath{$\nabla$} }\right\}
 \nonumber \\ \nonumber \\
                  & &   -\frac{e}{4M^2}\left\{\frac{}{}2 \kappa S(r)
                         \mbox{\boldmath{$\sigma$} }
                         \hspace{-.05in}\cdot\hspace{.05in}
                         \mbox{\boldmath{$\nabla$} }\times
                         \left[ {\bf A } \right]
                         \right.+ \mbox{\boldmath{$\sigma$} }
                         \hspace{-.05in}\cdot\hspace{.05in}{\bf A}
                         \times\mbox{\boldmath{$\nabla$}}
                         \left[ V(r)-S(r) \right]
 \nonumber\\ \nonumber\\
                  & &    \hspace{.75 in}\left.-
                         \left(\frac{}{} V(r)-S(r)
                         \frac{}{}\right)
                         \mbox{\boldmath{$\sigma$} }
                         \hspace{-.05in}\cdot
                         \hspace{.05in}\mbox{\boldmath{$\nabla$} }
                         \times\left[ {\bf A} \right]\right\}  .
\label{1-2-34b}
\end{eqnarray}
The calculations of the amplitudes proceed in the same manner
discussed in section \ref{Pauli}.
In particular the amplitude has the same form as that of equation
(\ref{non1-rel_s_mat}), but now the interaction Hamiltonian
is that of equation (\ref{1-2-34b}) while the wave functions are
solutions of a Schr\"{o}dinger-like equation using the
central and spin-orbit potentials of equation (\ref{1-2-34}).
Note that at first order the FW formalism
does not produce an acceptable nucleon wave function since there is
only a central potential and no spin-orbit potential, see equation
(\ref{1-2-34}).
The spin-orbit potential is borrowed from the second order and used
with the first order terms to get nucleon wave functions for use in
the first order which at least have a reasonable form.

\subsection{Differences Between the FW and Pauli Reduction Schemes}

The Pauli reduction scheme begins with the relativistic distorted
wave amplitude in which the initial and final nucleons are described
by Dirac wave functions.
The Dirac wave functions are solutions of the Dirac wave
equation (\ref{1.2}) with different strong potentials for
bound and continuum nucleons.

The amplitude resulting from the Pauli reduction procedure has
bound and continuum nucleons described by
nonrelativistic wave functions. These wave functions
are solutions of the Schr\"{o}dinger-like equation
(\ref{1.12}) with
central and spin-orbit potentials which in turn, are functions of
the Dirac potentials $S(r)$ and $V(r)$.

In the FW scheme the initial bound and final continuum nucleons
are described by the same Hamiltonian.
Due to the unrealistic potentials
used to describe the continuum nucleon,
the FW calculations provide a toy model which will shed more light
on the differences between the relativistic and nonrelativistic
calculations.
We should stress that unlike the Pauli reduction in
which the wave functions describing nucleons are the same in
all the calculations of different orders,
in the FW transformation the nucleon
wave functions are changed in each order of calculation due
to the contribution
of different terms to the wave equation.
Differences between these two reduction formalisms in
the free nucleon limit has been studied by Fearing {\it et al.}
\cite{FPS94}.
They find that for the interaction of a real photon with
a free particle at first order in the coupling constant
the FW and Pauli reductions produce the same nonrelativistic
interaction Hamiltonian.

\subsubsection{Differences between the Pauli and FW Hamiltonians}

In order to make instructive comparisons between the Pauli
and FW Hamiltonians we need to simplify the Hamiltonians
obtained through the Pauli reduction scheme of section
\ref{Pauli}.
First the potentials describing the initial and final
nucleons are taken to be the same;
in particular they will be the real
Hartree potentials used to describe the bound state
in the relativistic calculations. In addition, the
energies of the bound and continuum nucleons are set
equal to the nucleon mass in the Pauli interaction Hamiltonians.
In this limit the first order terms in the FW and Pauli
interaction Hamiltonians are exactly the same.
Differences will appear in terms of order $\frac{1}{M^2}$ and higher.
The difference between the interaction terms of the two schemes
up to second order, with the above modifications of
$H_I^{\mbox{\scriptsize Pauli}}$, is
\begin{eqnarray}
   H_I^{\mbox{\scriptsize FW}}- H_I^{\mbox{\scriptsize Pauli}}
               &=& \frac{e\omega}{8M^2}
                   \left\{  \mbox{\boldmath{$\sigma$} }
                   \hspace{-.05in}\cdot\hspace{.05in}
                   (\mbox{\boldmath{$\nabla$} }\times
                   \mbox{\boldmath{$A$} })-2
                   \mbox{\boldmath{$\sigma$} }\hspace{-.05in}
                   \cdot\hspace{.05in}({\bf A}\times
                   \mbox{\boldmath{$\nabla$} })\right\}
 \nonumber \\
               & &  -\frac{e\kappa }{4M^2}
                    \left\{ \left[ S(r)+V(r) \right]
                    \mbox{\boldmath{$\sigma$} }\hspace{-.05in}
                    \cdot\hspace{.05in}(\mbox{\boldmath{$\nabla$}}
                    \times {\bf A})\right\}  .
\label{1-2-42}
\end{eqnarray}
In the special case describing the interaction
of a photon with a free nucleon,
i.e in the limit when the strong potentials $S(r)$ and $V(r)$ are
set equal to zero, equation (\ref{1-2-42}) agrees
with the results of Fearing {\it et al.} \cite{FPS94}.
Detailed calculations show that the first term on the right
hand side of equation (\ref{1-2-42}) is very small for photon
energies less than a few hundred MeV, but becomes more
important at higher energies.
The second term
involves the sum of scalar and vector potentials.
At the origin the sum of these potentials is about -100 MeV.
Note that the square of the nucleon mass ($\sim$ 1000 MeV)
appears in the denominator of this term,
so it has a very small coefficient, with the result that
the second term on the right hand side of the above equation
also makes only a small contribution to the transition amplitude.
Thus the differences between the FW and Pauli interaction
Hamiltonians appear to be small if we restrict ourselves to second
order in the inverse nucleon mass.

The wave functions describing nucleons in the Pauli scheme are
the same for all orders of calculations while in the FW scheme
the wave functions change at every order.
The FW central potential at second order is similar in shape and
energy dependence to the central potential obtained via the
Pauli formalism.
A very interesting point to note is that there
is no spin-orbit potential at first order in the FW
formalism: it appears at second order with a shape similar to
that obtained in the Pauli scheme, but its magnitude is roughly
a factor of two smaller than the spin-orbit
potential arising in the Pauli scheme.

\subsection{Results of the FW Reduction}

In this section we compare the theoretical results of the
relativistic approach with those of the nonrelativistic
amplitudes obtained through FW reduction.
The results are shown for three different incident
photon energies.
As in the Pauli discussion of section \ref{Pauli-res}
we present three different types of nonrelativistic calculations,
namely: first order, medium-uncorrected second order
and medium-corrected second order.
Recall that in the first order nonrelativistic calculations
the wave functions are solutions of equation (\ref{1-2-28a}),
ignoring all the second order terms in $H_{0}$ (equation
(\ref{1-2-28})) except for the spin-orbit
potential.
In the second order calculations the wave functions are obtained
from equation (\ref{1-2-28a}) with all the first and second order
termsin $H_{0}$ included.
The Hartree potentials used in all calculations (relativistic
and nonrelativistic) are from reference \cite{HS}.
The graphs discussed in this section are labeled as in the Pauli
discussion of section \ref{Pauli-res} except for an obvious
change of notation.

Figure 3 shows the calculated observables for the
$^{16}O~(\gamma,p)~^{15}N$ reaction for a photon of energy
$E_\gamma =100 $ MeV.
Figure 3(a) shows the cross sections.
At small angles the first order nonrelativistic calculations
(dotted curve) are about an order of magnitude lower than the
relativistic calculations (solid curve), while for large angles
the first order calculations lie above the relativistic calculations.
Medium-uncorrected second order calculations (dot-dashed curve)
show substantial increase in the the magnitude of the cross sections
at small scattering angles as well as some change in the shape of
the resulting curve.
Medium-corrected second order calculations (dashed curve)
produce a noticeable change in the cross sections at backward angles.

The photon asymmetry calculations of Fig. 3(b) also show noticeable
differences between the first order nonrelativistic and
relativistic calculations at backward angles.
Medium-uncorrected second order calculations produce a change in
the magnitude and the shape of the asymmetry for scattering angles
greater than $80^\circ$.
Medium-corrected second order calculations produce a shift towards
larger angles resulting in a qualitatively similar shape to that
of the relativistic calculations.

Figure 4 shows the observables for the same reaction as Fig. 3 but
the photon energy in this case is $E_\gamma = 196$ MeV.
The cross section results are shown in Fig. 4(a), where we note
that the first order nonrelativistic calculations are generally
lower than the relativistic calculations by one to two orders of
magnitude.
They also fail to reproduce the dip near mid-angles.
Second order calculations lead to a drastic change in the cross
section with large differences due to medium corrections at both
forward and backward angles.
Medium-corrected second order calculations are in noticeably
closer agreement with relativistic calculations compared to the
medium-uncorrected ones.

Similar features are observed for the photon asymmetry calculations
(Fig. 4(b)).
Here again we notice large differences between the relativistic and
first-order nonrelativistic calculations.
Large differences also exist between medium-corrected and
medium-uncorrected calculations.
The level of agreement between the second order medium-corrected
calculations and the relativistic calculations is not the same as
observed in the case of cross sections.

The calculations shown in Fig. 5 for a photon energy of
$E_\gamma =312$ MeV show essentially the same qualitative features.

One characteristic that emerges from the above discussion is
that the full second order calculations
(medium-corrected calculations) in the FW scheme are not as close
to the relativistic
results as in the Pauli reduction case at the same photon energy.
This brings out an essential difference
between the Pauli and FW calculations:
The wave functions in the Pauli formalism remain the same while
different orders of the amplitude result solely from the expansion
of the interaction Hamiltonian.
The FW calculations, on the other hand, involve an expansion
affecting both the wave functions and the interaction
Hamiltonian simultaneously.
This difference is at the
root of the different convergence properties of the two approaches.
We find that in most cases, by second order the Pauli expansion is
quite close to the fully relativistic calculations,
provided medium corrections are taken into account.
In the FW scheme the level of agreement at the corresponding order
is inferior, indicating that the convergence in this scheme is
much slower than in the Pauli case.

\section{Conclusions}
\label{conclusion}

We have described two different nonrelativistic reduction
schemes
of the relativistic amplitude describing the knock-out
contribution to $(\gamma,p)$ reactions.
These reductions allow us to carry out controlled
comparisons between the relativistic and nonrelativistic
calculations of the reaction observables.
In the Pauli formalism the relativistic S matrix
is written in terms of nonrelativistic two-component
wave functions and an effective interaction Hamiltonian.
The effective Hamiltonian is expanded in powers of
$1/\left( E + M \right)$, where $M$ is the nucleon mass
and $E$ is its total energy.
In the limit $E\rightarrow M$, the first order interaction
terms are exactly the same as those appearing in the usual
nonrelativistic amplitude.
The nonrelativistic wave functions in this scheme are solutions
of the Schr\"{o}dinger-like wave equation (\ref{1.12}).
Detailed comparisons between the relativistic and first-order
nonrelativistic predictions for the differential cross sections
and photon asymmetries show large differences between the two
types of calculations.
The inclusion of terms to second order in $1/M$ in the interaction
Hamiltonian, where medium corrections effected by the nuclear
potentials are left out, does not lead to any substantial
improvement in the agreement between the relativistic and
nonrelativistic calculations.
On the other hand the expansion scheme shows explicit dependence
in the second order terms on the nuclear potentials.
When these medium corrections are taken into account the
nonrelativistic calculations converge
close to the relativistic results.
This indicates that the essential difference between the
relativistic and traditional nonrelativistic amplitudes, is the
absence in the
latter of the medium modification of the interaction Hamiltonian
as a consequence of the presence of the strong vector and scalar
potentials.

These conclusions are further supported through an analysis
based on the Foldy-Wouthuysen transformation of
the relativistic Hamiltonian describing a photon interacting
with a nucleon embedded in the nuclear medium.
The nonrelativistic wave functions for the bound and continuum
nucleons are solutions of the wave equation obtained as a result
of the transformation.
The scheme leads to a nonrelativistic amplitude
calculated to the desired order in $1/M$.
We use these amplitudes to carry out comparisons between
relativistic and nonrelativistic calculations in the manner
described above for the Pauli scheme.
We find that the medium modifications in the second order
calculations are important and their inclusion leads in general
to better agreement with the relativistic calculations.
However the convergence is not as efficient at this order
as in the Pauli case.
The reasons for this can be understood in terms of the
formal differences between the structure of the
nonrelativistic amplitude obtained using this transformation
as compared to the Pauli reduction case.
The wave functions obtained through the FW reduction
are different at each order in $1/M$,
in contrast to the Pauli wave functions
which are unchanged for all orders (recall that in the Pauli
reduction only the interaction Hamiltonian is expanded).

The basic result of the present work is that standard
nonrelativistic calculations of the knock-out amplitude do not
properly take into account the strong medium modifications
of the interaction Hamiltonian.
We have clarified this point through a comparison based on
nonrelativistic reduction of the relativistic amplitude
using both the Pauli and Foldy-Wouthuysen
reduction schemes.

\newpage
\section*{Appendix A}
In this appendix we give the explicit form of the six functions
involving Clebsch-Gordan coefficients and radial integrals which
are introduced in the amplitude of equation (\ref{1.23}).
\begin{eqnarray}
  \nonumber
  I_{l,L,J,L_B}           &=&  (L_B,l;0,0\mid L,0)
                               \int{r^2drf_B(r)j_{l}
                               (k_\gamma r)f_{LJ}(r)}  ,
  \\ \nonumber \\ \nonumber
  P^{M_B,\mu,\nu}_{L_B+1,l,L} &=&  \sqrt{\frac{L_B+1}{2L_B+1}}
                               (L_B+1,1;M_B+\mu-\nu,\nu\mid L_B,
                               M_B+\mu)
 \\ \nonumber
                          & &  \times (L_B+1,l:M_B+\mu-\nu,0
                               \mid L,M_B+\mu-\nu)
                               (L,l;0,0\mid L_B+1,0)
 \\ \nonumber
                          & &  \times\int{f_{LJ}(r)j_{l}
                               (k_{\gamma} r)\left( \frac{df_B(r)}
                               {dr}-L_B\frac{f_B(r)}{r} \right)r^2dr}  ,
\\ \nonumber \\ \nonumber
{\cal P}^{M_B,\mu,\nu}_{L_B-1,l,L} &=& \sqrt{\frac{L_B}{2L_B+1}}
                          (L_B-1,1;M_B+\mu-\nu,\nu\mid L_B,M_B+\mu)
  \\ \nonumber
                               & & \times (L_B-1,l:M_B+\mu-\nu,0
                                   \mid L,M_B+\mu-\nu)
  \\ \nonumber
                               & & \times (L,l;0,0\mid L_B-1,0)
 \\ \nonumber
                               & & \times\int{f_{LJ}(r)j_{l}
                                   (k_{\gamma} r)\left( \frac{df_B(r)}{dr}
                                   +\frac{L_B+1}{r}f_B(r) \right)r^2dr}  ,
\\ \nonumber \\ \nonumber
  H^{M_B,\mu,\nu}_{L,J,L_B}  &=&  (L,1/2;M_B-\mu-\nu,\mu\mid J,M_B-\nu)
  \\ \nonumber
                        & &\times (L_B,1/2;M_B-\mu,\mu\mid J_B,M_B)  .
\\ \nonumber  \\ \nonumber
  {\cal H}^{M_B,\mu}_{L,J,L_B} &=& (L,1/2;M_B+\mu,\mu\mid J,M_B+2\mu)
\\ \nonumber
                                 & & \times(L_B,1/2;M_B+\mu,
                                     -\mu\mid J_B,M_B)  ,
\\ \nonumber  \\ \nonumber
  C^{\mu}_{l,L,J,L_B}    &=&  (L_B,1/2;M_B+\mu,-\mu \mid J_B,M_B)
  \\ \nonumber
                      & &  \times (L,1/2;M_B+\mu,\mu\mid J,M_B+2\mu)
  \\ \nonumber
                      & &  \times(L,l;M_B+\mu,0\mid L_B,M_B+\mu)  ,
\label{1.24}
\end{eqnarray}

\newpage

\section*{Figure Captions}

\noindent Fig. 1: Differential cross section (a), and photon
asymmetry (b) for the reaction
$^{16}O(\gamma,~p)^{15}N$ at $E_{\gamma} = 100$ MeV. \\
Solid curve - full relativistic calculations. \\
Dotted curve - nonrelativistic calculations using the first order
               Hamiltonian $ H_I^{(1)}$ of equation (\ref{1.19}). \\
Dot-dashed curve - second order nonrelativistic calculations
                   (neglecting the nuclear potentials in
                   $H_I^{(2)}$ from equation (\ref{1.19})).
                   These are referred to as medium-uncorrected
                   second order calculations in the text. \\
Dashed curve - second order nonrelativistic calculations using the
               full $H_I^{(2)}$.
               These are referred to as medium-corrected second
               order calculations.
\\ \\
\noindent Fig. 2: Differential cross section (a), and photon
asymmetry (b) for the reaction of Fig. 1 but with
$E_{\gamma} = 312$ MeV.
Curves labelled as in Fig. 1.
\\ \\
\noindent Fig. 3: Differential cross section (a), and photon
asymmetry (b) for the reaction
$^{16}O(\gamma,~p)^{15}N$ at $E_{\gamma} = 100$ MeV. \\
Solid curve - full relativistic calculations. \\
Dotted curve - nonrelativistic calculations using the first order
               FW Hamiltonian $ H_I^{(1)}$ of equation (\ref{1-2-29})
               and nucleon wave functions obtained from equation
               (\ref{1-2-28a}) using the first order and
               spin-orbit terms of equation (\ref{1-2-28}). \\
Dot-dashed curve - second order nonrelativistic FW calculations
                   (neglecting the nuclear potentials in
                   ${H_I^{(2)}}^{\prime}$ from equation
                   (\ref{1-2-34b})),
                   and full second order potentials to
                   generate the wave functions.
                   These are referred to as medium-uncorrected
                   second order calculations in the text. \\
Dashed curve - second order nonrelativistic FW calculations using the
               full ${H_I^{(2)}}^{\prime}$.
               These are referred to as medium-corrected second
               order calculations.
\\ \\
\noindent Fig. 4: Differential cross section (a), and photon
asymmetry (b) for the reaction of Fig 3 but with
$E_{\gamma} = 196$ MeV.
Curves labelled as in Fig. 3.
\\ \\
\noindent Fig. 5: Differential cross section (a), and photon
asymmetry (b) for the reaction of Fig 3 but with
$E_{\gamma} = 312$ MeV.
Curves labelled as in Fig. 3.

%\end{document}

\begin{figure}
\begin{picture}(1100,400)(0,0)
\includegraphics{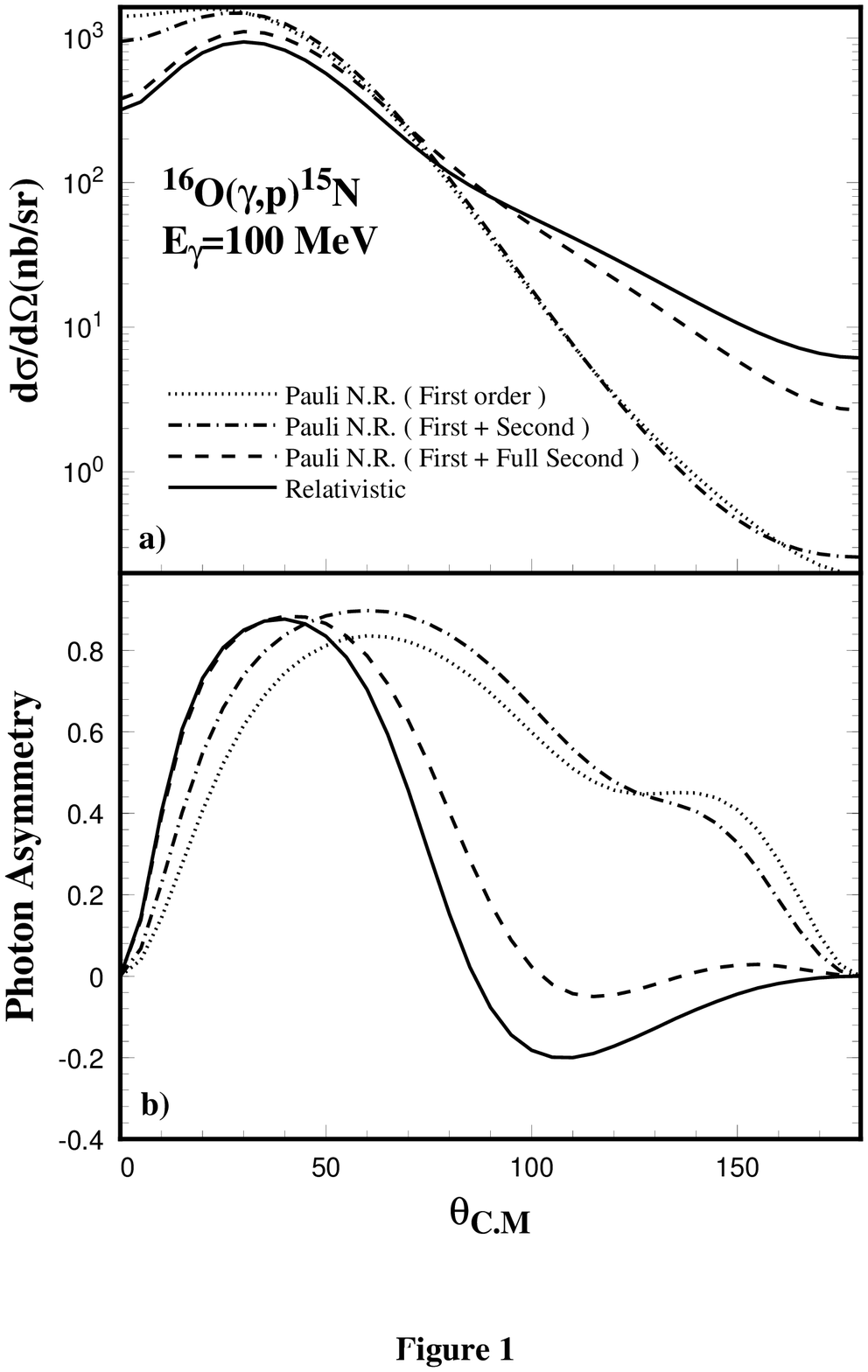}
\end{picture}
%\caption{}
\end{figure}

\begin{figure}
\begin{picture}(1100,400)(0,0)
\includegraphics{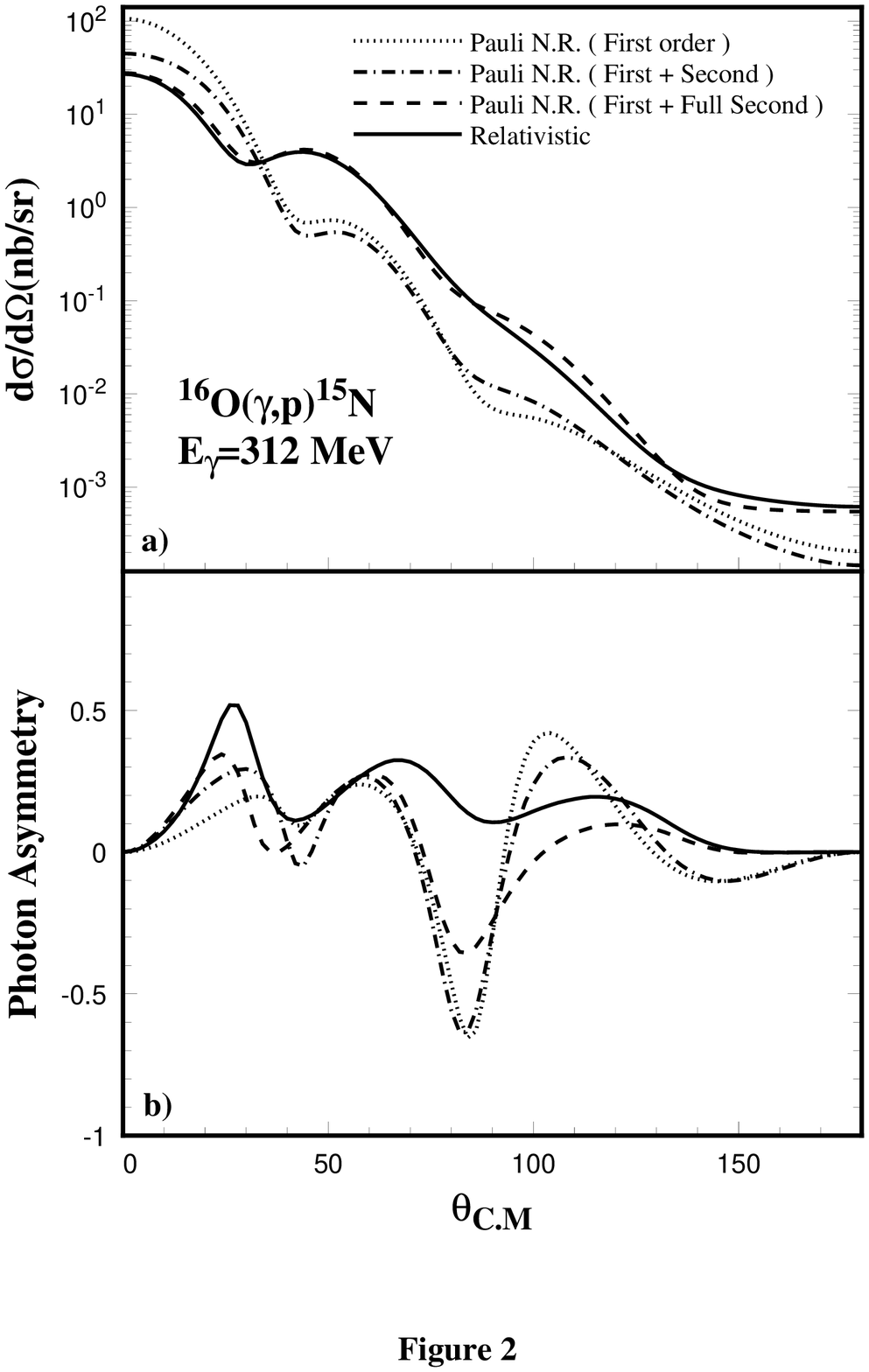}
\end{picture}
%\caption{}
\end{figure}

\begin{figure}
\begin{picture}(1100,400)(0,0)
\includegraphics{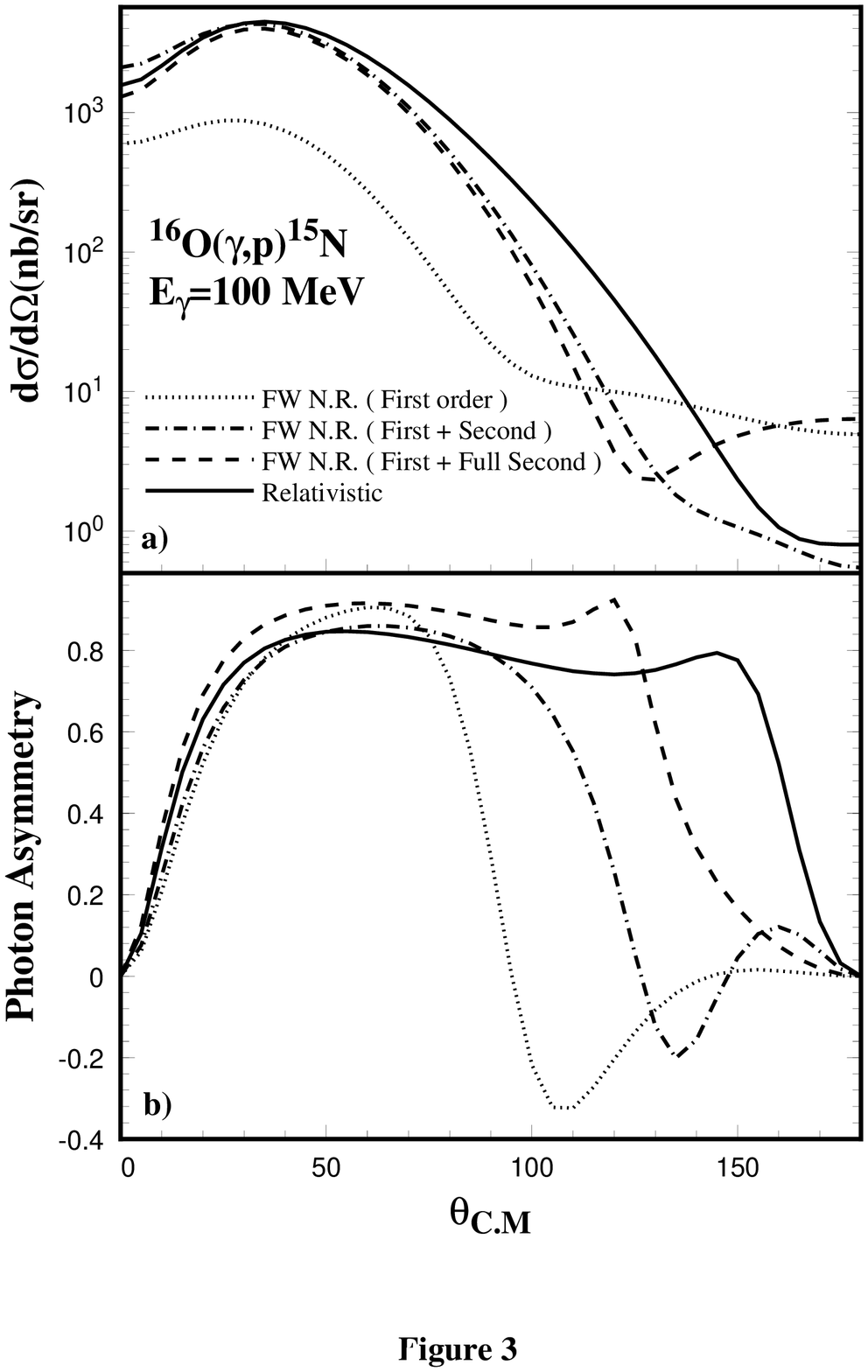}
\end{picture}
%\caption{}
\end{figure}

\begin{figure}
\begin{picture}(1100,400)(0,0)
\includegraphics{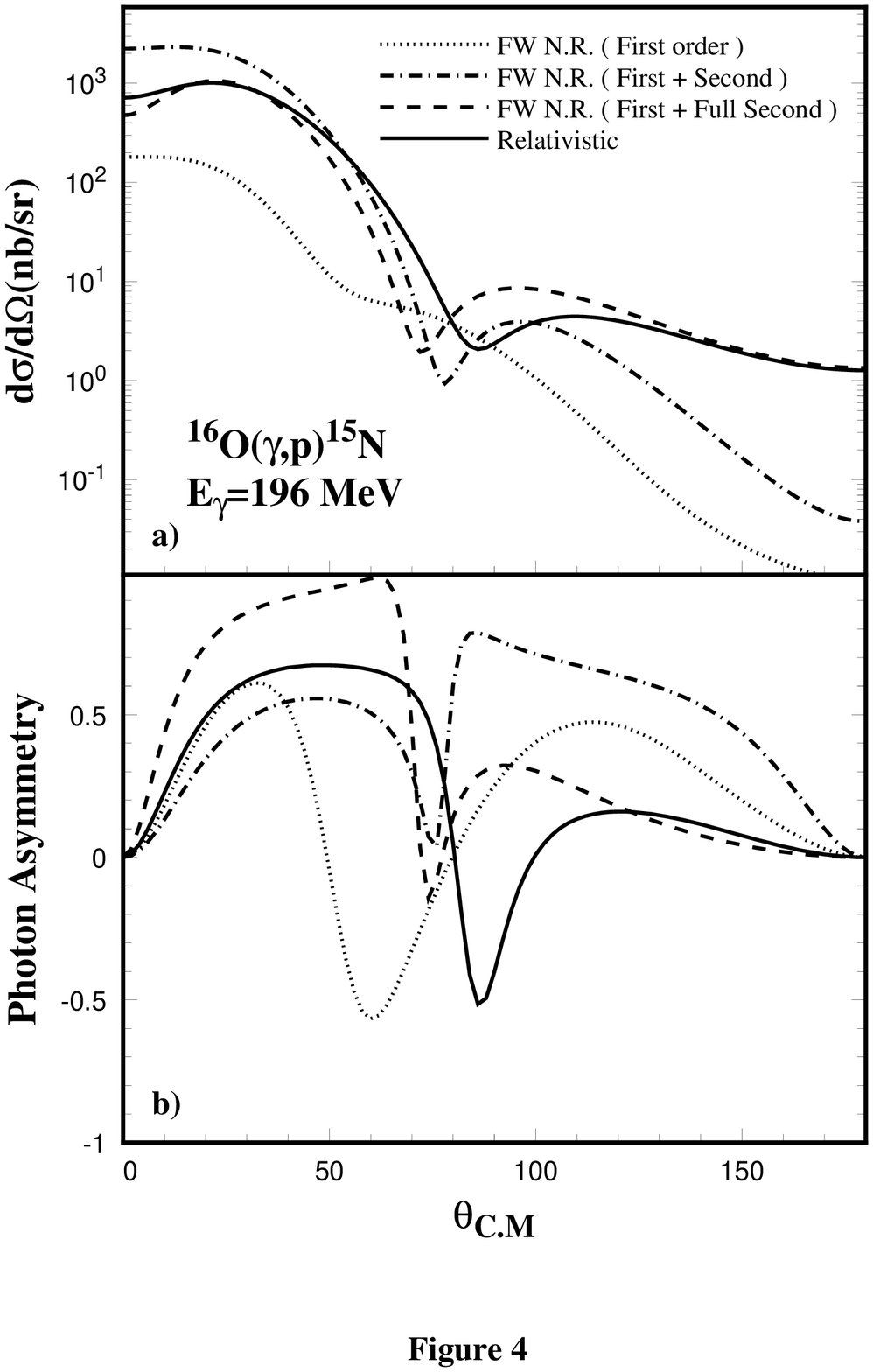}
\end{picture}
%\caption{}
\end{figure}

\begin{figure}
\begin{picture}(1100,400)(0,0)
\includegraphics{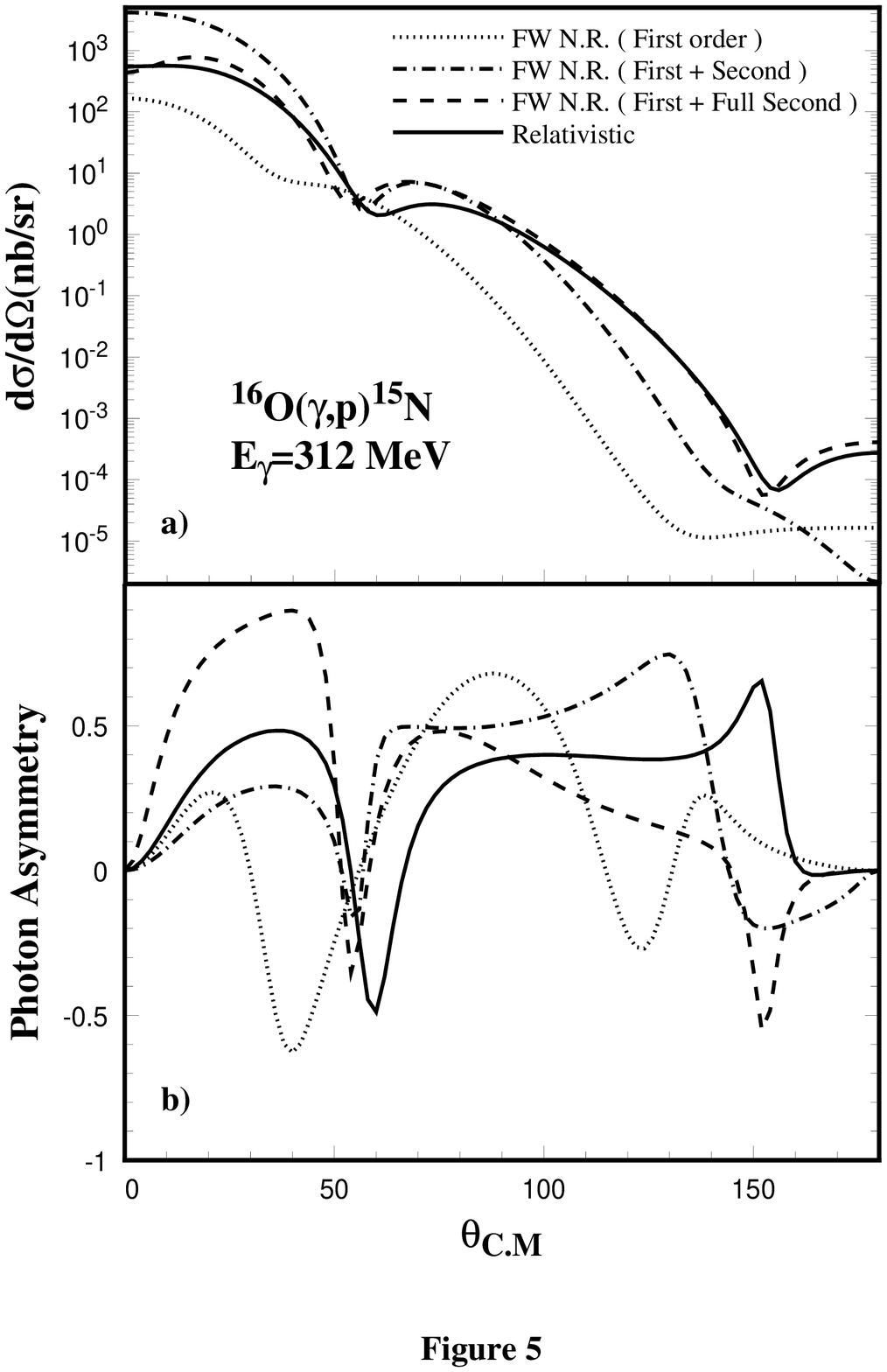}
\end{picture}
%\caption{}
\end{figure}


\begin{thebibliography}{99}

\bibitem {LS88} G.M. Lotz and H.S. Sherif,
                Phys. Lett. {\bf B210} (1988) 45; and
                Nucl. Phys. {\bf A537} (1992) 285.
\bibitem {Mc88} J.P. McDermott, E. Rost, J.R. Shepard and C.Y. Cheung,
                Phys. Rev. Lett. {\bf 61} (1988) 814.
\bibitem {TU85} R.S. Turley, E.R. Kinney, J.L. Matthews, W.W. Sapp,
                E.J. Scheidker, R.A. Schumacher, S.A. Wood, G.S. Adams,
                and R.O. Owens,
                Phys. Lett. {\bf 157B} (1985) 19.
\bibitem {FW}   L.L. Foldy and S.A. Wouthuysen,
                Phys. Rev. {\bf 78} (1950) 29.
\bibitem {MVVH} K.W. Mcvoy and L.Van Hove,
                Phys. Rev. {\bf 125} (1962) 1034.
\bibitem {HS94} M. Hedayati-Poor and H.S. Sherif,
                Phys. Lett. {\bf B326} (1994) 9.
\bibitem {HJS94}M. Hedayati-Poor, J.I. Johansson and H.S. Sherif,
                Phys. Rev. C {\bf 51} (1995) 2044.
\bibitem {NK87} H.W.L. Naus and J.H. Koch, Phys. Rev. C {\bf 36}
                (1987) 2459.
\bibitem {BDQM} J.D. Bjorken and S.D. Drell,
                {\em Relativistic Quantum Mechanics},
                (McGraw-Hill Book Company 1964).
\bibitem{SW}    B.D. Serot and J.D. Walecka, Advances in Nuclear
                Physics (J.W. Negele and E. Vogt,eds.) Vol.{\bf 16},
		Plenum Press, New York (1986)
\bibitem {EC}   E.D. Cooper, S. Hama, B.C. Clark and R.L. Mercer,
                Phys. Rev. C {\bf 47} (1993) 297.
                (1981) 91.
\bibitem {HS}   C.J. Horowitz and B.D. Serot,
                Nucl. Phys. {\bf A368} (1986) 503.
\bibitem {CHM}  B.C. Clark, S. Hama and R.L. Mercer,
                AIP conf. proc. N. {\bf 97}, ed:
                H.O. Mayer (1982) 260;
                J. Raynal, Aust. J. Phys. {\bf 43} (1990) 9;
                G.Q. Li, J. Phys. {\bf G19} (1993) 1841.
\bibitem {COMAC83}E.D. Cooper, A.O. Gattone and M.H. Macfarlane,
                Phys. Lett. {\bf B130} (1983) 359.
\bibitem {GGZKRS} V.B. Ganenko, V.A. Gushchin, Yu. V. Zhebrovskii,
                L. Ya. Kolesnikov, A.L. Rubashkin and P.V. Sorokin,
                JETP Lett. {\bf 47} (1988) 519;
                G.S. Blanpied {\it et al.},
                Phys. Rev. Lett. {\bf 67} (1991) 1206.
\bibitem {GE}   Gerhard Martin Lotz,
                Ph.D. thesis, University Of Alberta, 1989.
\bibitem {BGP81} S. Boffi, C. Giusti and F.D. Pacati,
                Nucl. Phys. {\bf A359}
\bibitem{FPS94} H.W. Fearing, G.I. Poulis and S. Scherer, Nuc. Phys.
                {\bf A570},  (1994) 657.

\end{thebibliography}
\end{document}